\documentclass[twocolumn,showpacs,amsmath,amssymb,superscriptaddress]{revtex4}
\usepackage{graphicx}
\usepackage{dcolumn}
\usepackage{bm}
\usepackage[latin1]{inputenc}
\usepackage{float}
\usepackage{bbold}

\begin{document}

\title{Two electron entanglement in quasi-one dimensional system: Role of resonances}

\author{Alexander L\'opez, Otto Rend\'on, V\'\i ctor M. Villalba}
\affiliation{Centro de Física, Instituto Venezolano de Investigaciones Científicas.
IVIC, Apartado 21827,Caracas 1020 A, Venezuela.}
\author{Ernesto Medina}
\affiliation{Centro de Física, Instituto Venezolano de Investigaciones Científicas.
IVIC, Apartado 21827,Caracas 1020 A, Venezuela.}
\affiliation{Physics Department, Boston University, Boston, Massachusetts 02215}

\date{\today}

\begin{abstract}
We analyze the role of resonances in two-fermion entanglement production for a quasi one-dimensional two channel scattering problem. We solve exactly for the problem of a two-fermion antisymmetric product state scattering off a double delta well potential. It is shown that the two-particle concurrence of the post-selected state has an oscillatory behavior where the concurrence vanishes at the values of momenta for virtual bound states in the double well. These concurrence zeros are interpreted in terms of the uncertainty in the knowledge of the state of the one particle subspace reduced one particle density matrix. Our results suggest manipulation of fermion entanglement production through the resonance structure of quantum dots.
\end{abstract}
\maketitle

Entanglement production and quantification has been given much recent attention due to its importance as a resource for quantum information and quantum communication\cite{Reviews,nielsen}. In this direction, there have been recent proposals for producing bipartite fermionic entangled states in the solid state environment focusing on the role of direct interaction between particles. Some of these approaches involve direct coulomb interactions in quantum dots \cite{burkard1} and interference effects\cite{Yamamoto}, phonon mediated interactions in superconductors \cite{Superconductors,Samuelsson} and Kondo-like scattering of conduction electrons\cite{Bose}. Nevertheless, it has been shown that fermion entanglement can be achieved in absence of such interactions\cite{BeenakkerTeleport} in the form of particle hole entanglement even when fermions are injected from thermal reservoirs. In such a setup the orbital degree of freedom is entangled. Other implementations based on the non-interacting scheme have been proposed that entangle the spin degree of freedom and are thus more robust to decoherence \cite{LebedevSpinEnt} because of the weaker coupling of the spin to the environment.

\begin{figure}
\includegraphics[width=8cm]{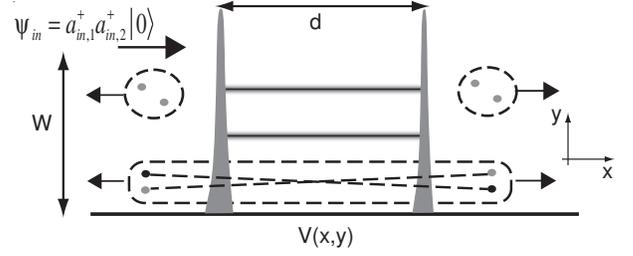}
\caption{\label{fig1} {The scattering setup for a quasi-one dimensional wire of width $w$ and a
scattering region with potential $V(x,y)$ consisting of a sequence of two delta potentials separated by a distance $d$ . A two electron antisymmetrized wave function is injected at the left. The outgoing products according to Eq.\ref{waveout} consist of three terms: a) two electrons are reflected b) two electrons are transmitted and c) one electron is transmitted and the other reflected.}}
\end{figure}

In this work we address the problem of entanglement generation for electrons in the context of a two channel quasi-one dimensional conductor\cite{mello}, following the scattering matrix formalism of reference\cite{BeenakkerTeleport}. For the scattering region, we choose a double delta potential, separated a distance $d$. Such a potential is the simplest, that exhibits resonances, that can be analytically handled. The problem is solved for the concurrence\cite{wootters,wootters2} exactly, for all values of the barrier heights and separation as a function of the incoming electron momenta. The concurrence of the entangled post selected state is found to oscillate while its envelope decays as a function of electron momentum ($k_i$) difference $\Delta k= k_2-k_1$. We find that the concurrence is exactly zero when one or both of the $k$ values hits the resonant states for the potential well. The concurrence zeros are then interpreted in terms of the uncertainty of the state in the one particle subspace by obtaining the reduced density matrix. We thus determine the role of resonances on the entangling properties of the well demonstrating new possibilities for fermion entanglement control.
We consider in detail the independent channel scenario but quantitative changes due to channel mixing will be briefly discussed. Here we ignore the effects of temperature since they have been assessed in a general way in reference\cite{FiniteTemp} and will not change qualitatively the results reported here if a critical temperature is not reached.
The set up for the system considered is depicted in Fig. \ref{fig1}  where a two electron wavefunction is injected at the left of a quasi-one dimensional conductor wire. The electrons move freely until they enter the interacting domain with potential $V(x,y)$. Each electron is considered to pertain to a separate channel in the incoming lead  and gets transmitted or reflected within the same two channels.

Let us set up the problem in a first quantized description. Schr$\ddot{\rm o}$dinger's equation is given by
\begin{equation}
\left [-\frac{\hbar^2}{2m}\left (\frac{\partial^2}{\partial x^2}+\frac{\partial^2}{\partial y^2}\right )+V(x,y)\right ]\psi(x,y)=E\psi(x,y).
\end{equation}
The potential $V(x,y)$ acts in a finite region of the coordinate $x$ (see Fig. \ref{fig1}). The boundary conditions on the wire are such that $\psi(x,0)=\psi(x,w)=0$. In the free regions (the leads) $V(x,y)=0$ and using the definitions $k^2=2mE/\hbar^2$ and $U(x,y)=2mV(x,y)/\hbar^2$, we can write the Schr$\ddot{\rm o}$dinger equation as
\begin{equation}
\left [\frac{\partial^2}{\partial x^2}+\frac{\partial^2}{\partial y^2}+k_{\|}^2 +K_{\bot,n}^2 \right ]\phi(E_{\|},x)\chi_n(y)=0,
\end{equation}
where $k^2=k_{\|}^2+K_{\bot,n}^2$. The eigenfunction $\phi(E_{\|},x)$ is given by
${e^{isk_{\|}x}}/{\sqrt{2\pi\hbar^2k_{\|}/m}}$,
$K_{\bot,n}=n\pi/w$ and $\chi_n(y)=\sqrt{\frac{2}{w}}\sin K_{\bot,n}y$. The integer $n$ denotes the channel number. When $k^2>K_{\bot,2}^2$ both channels are open.

In the potential region $\psi(x,y)=\sum_{n=1}\psi_n(x)\chi_n(y)$ where, if we define
$U_{mn}(x)=\int_0^w\chi_n(y)U(x,y)\chi_m(y)dy$,
we get the system of coupled equations
\begin{equation}
\left [\frac{\partial^2}{\partial x^2}+k^2-K_{\bot,n}^2 \right ]\psi_n(x)=\sum_{m=1}U_{mn}(x)\psi_m(x).
\end{equation}
We denote the difference $k^2-K_{\bot,n}^2=k_n^2$, omitting the suffix $\|$, and we always understand that the difference is positive.

We now fix $n=1,2$ and choose $U_{mn}(x)=u_{mn}v(x)$, with $v(x)=\delta(x-d/2)+\delta(x+d/2)$ and $u_{mn}=u^*_{nm}$. To begin with, we take $u_{12}=u_{21}=0$ which means no channel mixing.
The composition of two delta scatterers at $x=-d/2$ and $x=d/2$, in series, with corresponding scattering matrices $S_I$ and $S_{II}$, gives the symmetric $S$ matrix
\begin{eqnarray}
S=\left (\begin{array}{cc}
r_I+t'_Ir_{II}\frac{1}{\mathbb{1}-r'_Ir_II}t_I & t'_I\frac{1}{\mathbb{1}-r'_Ir_{II}}t'_{II} \\
t'_I\frac{1}{\mathbb{1}-r'_Ir_{II}}t'_{II} & r_{II}+t_{II}r'_I\frac{1}{\mathbb{1}-r_{II}r'_I}t'_{II}
\end{array}\right ),
\end{eqnarray}
where $\mathbb{1}$ is the $2\times 2$ identity matrix.

We can now  use Beenakker's\cite{BeenakkerTeleport} approach to arrive at the expression for the output wavefunction. Using the same notation
\begin{equation}
|\Psi_{in}\rangle =a^{\dagger}_{in,1}(\epsilon)a^{\dagger}_{in,2}(\epsilon)|0\rangle,
\end{equation}
where $a^{\dagger}_{in,l}$ creates one electron on the left in channel $l$. Now $b^{\dagger}_{in,j}(\epsilon)$ creates an electron in channel $j$ incident from the right so that in matrix notation the input state can be written as
\begin{displaymath}
|\Psi_{in}\rangle = {a^{\dagger}_{in} \choose  b^{\dagger}_{in}}
\left (\begin{array}{cc}
\frac{i}{2}\sigma_y & 0\\
0 & 0
\end{array}\right )
{a^{\dagger}_{in} \choose b^{\dagger}_{in}}|0\rangle,
\end{displaymath}
where the vectors are $4\times1$ and the matrix is $4\times4$, because there are two channel indices on the right and the left. The relation between input and output channels is given by the scattering matrix 
\begin{displaymath}
{a_{out} \choose b_{out}}=\left (\begin{array}{cc}
r & t' \\
t & r'
\end{array}\right ){a_{in} \choose b_{in}}.
\end{displaymath}
The entries $r, t, r'$ and $t'$ are $2\times2$ reflection and transmission matrices. After some algebra one arrives at the exact relation
\begin{eqnarray}
\label{waveout}
|\Psi_{out}\rangle & = & (a^{\dagger}_{out}r\sigma_y t^{T}b^{\dagger}_{out}+[r\sigma_yr^{T}]_{12}
a^{\dagger}_{out,1}a^{\dagger}_{out,2}+\nonumber \\
& & [t\sigma_yt^{T}]_{12} b^{\dagger}_{out,1}b^{\dagger}_{out,2}) |0\rangle .
\end{eqnarray}
For no channel mixing, and in terms of our particular potential, the $r$ and $t$ matrices are given through
\begin{displaymath}
 r_{jj}=r_je^{-ik_jd}\left(1+\frac{t_j^2e^{2ik_jd}}{1-r_j^2e^{2ik_jd}}\right), \quad t_{jj}=\frac{t_j^2}{1-r_j^2e^{2ik_jd}},
\end{displaymath} 
and $r_{12}=r_{21}=t_{12}=t_{21}=0$, where the indices of the reflection and transmission amplitudes refer to the channel, and $r_j$, $t_j$ ($j=1,2$) have the expressions $r_j=(u_{jj}/{2ik_j})/({1-{u_{jj}}/{2ik_j}})$ and  $t_j={1}/({1-{u_{jj}}/{2ik_j}})$.
The new element here is that now we have energy dependent transmission and reflection amplitudes and the presence of resonances because of virtual states in the barrier. 

In order to derive from the scattering result of Eq.\ref{waveout} an entangled state one must now post-select or project out the appropriate component. In this case one can post-select by coincidence measurements where electrons are detected simultaneously at opposite branches of the double barrier, a well known experimental tool in optics\cite{Shih}. The useful term is first order in $t$ and $r$ generating particles on both sides of the double barrier
\begin{equation}\label{EntangledState}
|\Phi\rangle=\frac{1}{\sqrt{{\rm Tr}\gamma\gamma^{\dagger}}}a^{\dagger}_{out}\gamma b^{\dagger}_{out}|0\rangle,
\end{equation}
where $\gamma=r\sigma_y t^{T}$. The state is appropriately normalized. In order to compute the concurrence we use the convenient definition\cite{LossConcurrence}, that reduces the problem to identifying the ${\rm \bf W} $ matrix in the expansion
$|\Phi\rangle =\sum_{\alpha,\beta}{\rm W}_{\alpha\beta}~a^{\dagger}_\alpha b^{\dagger}_\beta|0\rangle$,
where $\alpha,\beta\in \{1,2,3,4\}$ and ${\rm W}_{\alpha\beta}$ can be assumed antisymmetric. Expanding the product $a^{\dagger}_{out}\gamma b^{\dagger}_{out}|0\rangle$ one finds
\begin{eqnarray}
a^{\dagger}_{out}\gamma b^{\dagger}_{out}|0\rangle & = & [\gamma_{11}a^{\dagger}_{out,1} b^{\dagger}_{out,1} +\gamma_{21} a^{\dagger}_{out,2} b^{\dagger}_{out,1}+
\nonumber \\
& & \gamma_{12}a^{\dagger}_{out,1}b^{\dagger}_{out,2}+ \gamma_{22}a^{\dagger}_{out,2}b^{\dagger}_{out,2}]|0\rangle.\nonumber 
\end{eqnarray}
Then the antisymmetric part of ${\rm W}_{\alpha\beta}$ is given by
\begin{displaymath}
{\rm\bf W}=\frac{1}{2\sqrt{{\rm Tr}\gamma\gamma^{\dagger}}}\left (\begin{array}{cccc}
0 & 0 & \gamma_{11} & \gamma_{12} \\
0 & 0 & \gamma_{21} & \gamma_{22} \\
-\gamma_{11} & -\gamma_{21} & 0 & 0 \\
-\gamma_{12} & -\gamma_{22} & 0 & 0 \\
\end{array}\right ).
\end{displaymath}
\begin{figure}
\includegraphics[width=7cm]{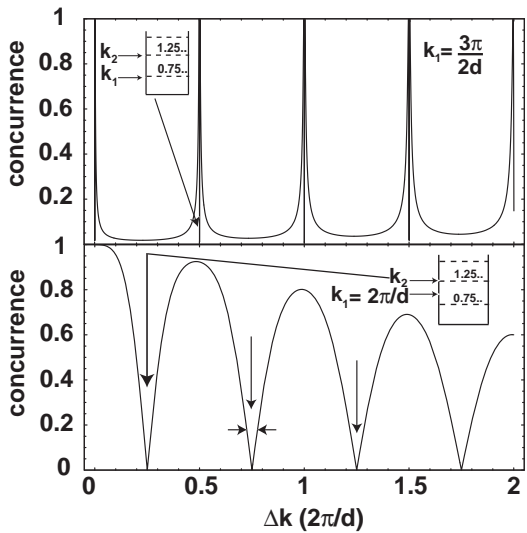}
\caption{\label{fig2} {Concurrence as a function of the wavenumber
difference $\Delta k=k_2-k_1$ (in $2\pi/d$ units) without channel mixing. The barrier heights have been fixed to $u_0=(2\pi/d)\cdot(1/100)$ and $k_1,k_2$ take values as shown. In the bottom panel the concurrence is zero whenever $k_2$ hits a resonance while $k_1$ is in between resonances as indicated in the inset. In the top panel now $k_1$ is in the vicinity of a resonance and a concurrence maximum occurs when a $k_2$ closes onto another (going exactly to zero when the resonance is hit).}}
\end{figure}
\begin{figure}
\includegraphics[width=7cm]{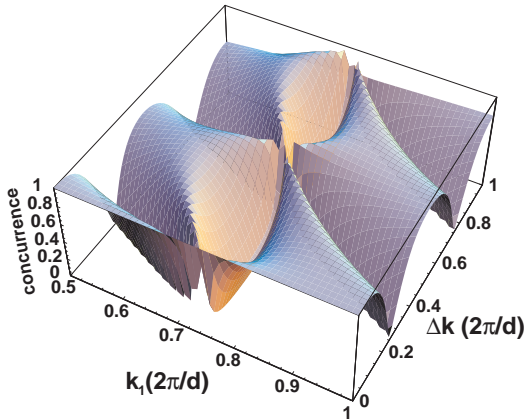}
\caption{\label{fig3} {(Color online only) Concurrence as a function of both $k_1$ and $\Delta k=k_2-k_1$. The strength of the delta potential has been fixed at $u_0=2\pi/d~(1/100)$. Limiting behaviors depicted in figure 2 connect smoothly as $k_1$ is varied. The zeros of concurrence always correspond to $k_2$ hitting a resonance.}}
\end{figure}

The expression for the concurrence is $\eta=|\langle {\tilde \Psi}|\Psi\rangle|=\varepsilon^{\alpha\beta\mu\nu}{\rm W}_{\alpha\beta}{\rm W}_{\mu\nu}$, where $\varepsilon^{\alpha\beta\mu\nu}$ is the totally antisymmetric unit tensor in 4 dimensions. Then $\eta=8|{\rm W}_{12}{\rm W}_{34}+{\rm W}_{13}{\rm W}_{42}+{\rm W}_{14}{\rm W}_{23}|$. Computing $\eta$ for the ${\rm W}_A$ matrix above gives $2|{\rm det}\gamma|/{\rm Tr}\gamma\gamma^{\dagger}$. For the general case including channel mixing the matrix $\gamma$ is given by
\begin{equation}\label{gama}
\gamma=\left (\begin{array}{cc}
r_{12}t_{11}-r_{11}t_{12} & r_{12}t_{21}-r_{11}t_{22} \\
r_{22}t_{11}-r_{21}t_{12} & r_{22}t_{21}-r_{21}t_{22}
\end{array}\right ).
\end{equation}
It is obvious that when there is no channel mixing the $\gamma$ matrix is anti-diagonal and the resulting concurrence is then
\begin{equation}\label{eta}
\eta=\frac{2|r_{22}||t_{11}||r_{11}||t_{22}|}{|r_{22}|^2|t_{11}|^2+|r_{11}|^2|t_{22}|^2}~  .
\end{equation}
Substituting the values for the reflection and transmission amplitudes one can derive $\eta$ as a function of the incoming wavevectors. Since both transmission and reflection matrices are $k$-dependent, the resonant properties of the device are relevant for the two particle concurrence. It is worth noting a crucial point: If one computes de concurrence without post-selecting the result is zero. This is consistent with the fact that local processes cannot change the entanglement, which for the input state is zero.

In the absence of channel mixing we depict, in Fig. \ref{fig2}, $\eta$ as a function of the difference in the magnitude of wavevectors $\Delta k=k_2-k_1$. We have scaled all wavevectors and the magnitude of the delta potential so they are in units of $2\pi/d$. Having fixed the height of the barriers and the distance between them, the one particle resonances occur at fixed values shown in the figure panels as insets in units of $2\pi/d$. Two well defined limiting behaviors occur: The bottom panel depicts the case where $k_1=2\pi/d$; here we notice that as a function of $\Delta k$, $\eta$ shows an oscillating and decreasing pattern. The minima, which are exactly zeros of the concurrence, occur when $k_2$ values hit a resonance while $k_1$ is in between resonances as indicated by the inset. Succesive zeros coincide with succesive one particle resonances.

The top panel of Fig.\ref{fig2}, shows a second behavior occuring when $k_1$ is in the vicinity of a resonance, as indicated by the panel inset. The concurrence then is only appreciable within the resonance width, and goes to zero, exactly, when $k_2$ wavevector hits resonances. The full range of behaviors described and their crossovers between that of the bottom panel and top panel in Fig.\ref{fig2}, are shown in Fig.\ref{fig3} in a representative range of $k_1$ and $\Delta k$.

A useful tool to gain intuition on entanglement is to obtain the reduced one particle density matrix of the state in Eq.\ref{EntangledState}. If there is entanglement, the resulting density matrix represents a mixed state showing there is uncertainty in the state of the particle. Vanishing of entanglement is then evidenced by the certainty of a particular state. We stress though, that there is no new information regarding entanglement that is not already assessed in Eq.\ref{eta}. Setting up the two particle density matrix $\rho_{2}=|\Phi\rangle \langle \Phi |$ and the tracing over one of the particles results in the matrix
\begin{eqnarray}\label{densitymatrix}
\rho_1={\rm Tr}_1\rho_2 &=& \frac{R_{22}^2T_{11}^2}{2{\rm Tr}\gamma\gamma^{\dagger}}\left (|{\mathcal R}_1\rangle\langle {\mathcal R}_1| + |{\mathcal L}_1\rangle\langle {\mathcal L}_1| \right )+\nonumber\\
& & \frac{R_{11}^2T_{22}^2}{2{\rm Tr}\gamma\gamma^{\dagger}}\left (|{\mathcal R}_2\rangle\langle {\mathcal R}_2| + |{\mathcal L}_2\rangle\langle {\mathcal L}_2| \right )
\end{eqnarray}
where $R_{ii}^2=|r_{ii}|^2$ and analogously for $T_{ii}$. The kets are defined as $b_{out,i}^{\dagger}|0\rangle=|{\mathcal R}_i\rangle$ and  $a_{out,i}^{\dagger}|0\rangle=|{\mathcal L}_i\rangle$. This is a mixed state (as can be seen by tracing over $\rho_1^2$) where the remaining electron is projected, with probability $R_{22}^2T_{11}^2/2{\rm Tr}\gamma\gamma^{\dagger}$, onto channel 1 on the right, $|{\mathcal R}_1\rangle$. Note that for arbitrary reflection and transmission probabilities, the electron can be in any of the two channel states (1 or 2) signaling entanglement between the two electrons. This indeterminacy is destroyed once we hit a single particle resonance so that $T_{11}=1$ (so that $R_{11}=0$) becoming a certainty since $\rho_1=1/2 (|{\mathcal R}_1\rangle\langle {\mathcal R}_1| + |{\mathcal L}_1\rangle\langle {\mathcal L}_1|)$ and the electron can only be in channel 1. This explains the zeros of the concurrence at the single particle resonances.

Some additional features of the figure can be accounted for using the above expression: The first maximum ($k_1=k_2$) in the bottom panel of Fig.\ref{fig2} corresponds to the maximum uncertainty in distinguishing one particle from the other. In this situation, the probability amplitudes for transmission (reflection) $t$ ($r$) through either channel are the same (see Eq. \ref{eta}). As the wavevector difference increases the concurrence envelope function drops monotonously indicating the uncertainty is also reduced. This can be seen from Eq.\ref{densitymatrix} by noting that as $\Delta k$ increases ($k_2$ increases) the corresponding transmission coefficient $T_{22}$ grows reducing the state uncertainty by the argument given for the resonances. The introduction of mixing terms, involved in $\gamma_{12},\gamma_{21}$, change the scenarios described above only quantitatively. The resonances will shift positions and the envelope of the concurrence as a function of $\Delta k$ will now be non-monotone. As $\eta$ depends on the determinant and the trace operations, one can diagonalize the new $\gamma$ matrix and formally use equivalent expressions to the ones above.

Although it is not the intent of the paper to propose a practical experimental setup to produce entanglement, resonance effects are ubiquitous for any quantum dot system coupled to external leads. The width of the resonances can be controlled by the coupling of the dot to the leads and the resonance position in energy can be adjusted, relative to the Fermi levels in the leads, by a gate voltage. The results of our paper show that the manipulation of the resonances will lead to the control of the degree of the entanglement. The realization of two separate electron channels in the same region has been addressed differently in the literature by tapping into edge states that can provide two quantum numbers\cite{BeenakkerTeleport} for the incoming electrons. A gated quantum dot can be placed in the vicinity of the edge states with a controlled coupling so as to modulate the resonance characteristics of the dot. The outgoing electrons can be detected by gate electrodes that mix channels
appropriately so as to change the measuring eigenbasis and compute for example
the Bell inequalities or other measures of quantum correlations. This setup for
the case of a spectrally structureless beam splitter has been described previously on the basis of current-current correlations in ref.\cite{Samuelsson}.

This work analized the role of resonances in entanglement production in a quasi one dimensional two electron system inspired in the electron-hole entangler of ref.\cite{BeenakkerTeleport}. Although we have restricted to elastic scattering (no channel mixing), we found that by tuning the channel momenta or, equivalently, the resonant levels of a double barrier (through for example gate voltages) the quantum correlations associated with the post-selected scattering process can be manipulated in a controlled fashion. Needless to say, the response to resonances of electron-electron correlations is relevant in the context of electron hole entanglement\cite{BeenakkerTeleport}. 


\begin{acknowledgments}
A.L. acknowledges L. Gonz\'alez for useful discussions. E. M. was supported by FONACIT through grant S3-2005000569.

\end{acknowledgments}


\begin{thebibliography}{}

\bibitem{Reviews} T. Martin, A. Crepieux, and N. Chtchelkatchev, in {\it Quantum Noise in Mesoscopic Physics}, Edited by Yu. V. Nazarov, NATO Science Series II. Vol. 97 (Kluwer, Dordrecht, 2003).

\bibitem{nielsen} M.A. Nielsen and I.L. Chuang,
\newblock {{\it Quantum Computation and Quantum Information,} {Cambridge University Press}, (2000).}

\bibitem{burkard1} G. Burkard, D. Loss, and E. V. Sukhorukov, Phys. Rev. B {\bf 61}, R16303 (2000); D. Loss and E. V. Sukhorukov, Phys. Rev. Lett. {\bf 84}, 1035 (2000); D. S. Saraga, and D. Loss, Phys. Rev. Lett. {\bf 90}, 166803 (2003).

\bibitem{Yamamoto} W. D. Oliver, F. Yamaguchi, and Y. Yamamoto, Phys. Rev. Lett. {\bf 88} 037901 (2002); G. Le\'on, O. Rend\'on, H. M. Pastawski, V. Mujica, and E. Medina, Europhysics Lett. {\bf 66}, 624 (2004).

\bibitem{Superconductors}
G. B. Lesovik, T. Martin, and G. Blatter, Eur. Phys. J. B {\bf 24}, 287 (2001); P. Recher, E. V. Sukhorukov, and D. Loss, Phys. Rev. B {\bf 63}, 165314 (2001); P. Recher and D. Loss, Phys. Rev. B {\bf 65}, 165327 (2002).

\bibitem{Samuelsson} P. Samuelsson, E.V. Sukhorukov and M. Buttiker, Phys. Rev. Lett. {\bf 91}, 157002 (2003).

\bibitem{Bose} A. T. Costa, and S. Bose, Phys. Rev. Lett. {\bf 87}, 277901 (2001).

\bibitem{BeenakkerTeleport} C.W.J. Beenakker, C. Emary, M. Kindermann, and J. L. van Velsen, Phys. Rev. Lett.  {\bf 91}, 147901 (2003).

\bibitem{LebedevSpinEnt} A. V. Lebedev, G. Blatter, C. W. J. Beenakker, and G. B. Lesovik, Phys. Rev. B {\bf 69}, 235312 (2004); A. Di Lorenzo and Yu. V. Nazarov, Phys. Rev. Lett. {\bf 94}, 210601 (2005).

\bibitem{mello} P.A. Mello and N. Kumar, {\it Quantum Transport in Mesoscopic systems: Complexity and Statistical Fluctuations} (Oxford University Press, 2004).

\bibitem{wootters} S. Hill and W.K. Wootters, Phys. Rev. Lett. {\bf 78}, 5022 (1997).

\bibitem{wootters2} W.K. Wootters, Phys. Rev Lett. {\bf 80}, 2245 (1998).

\bibitem{FiniteTemp} B. V. Fine, F. Mintert, and A. Buchleitner, Phys. Rev. B {\bf 71}, 153105 (2005); C. W. J. Beenakker, arXiv:cond-mat/0508488.

\bibitem{Shih} T. B. Pittman, and J. D. Franson, Phys. Rev. Lett. {\bf 90}, 240401 (2003).

\bibitem{LossConcurrence} J. Schliemann, D. Loss, and A.H. MacDonald, Phys. Rev. B {\bf 63}, 085311 (2001).

\end{thebibliography}
\end{document}